\documentstyle[amssymb,aps,preprint]{revtex}
\tightenlines

\begin{document}
\title{Thermal fluctuations correction to magnetization and specific heat of vortex
solids in type II superconductors.}
\author{Dingping Li\thanks{%
e-mail: lidp@mono1.math.nctu.edu.tw} and Baruch Rosenstein\thanks{%
e-mail: baruch@vortex1.ep.nctu.edu.tw}}
\address{{\it National Center for Theoretical Sciences and} \\
{\it Electrophysics Department, National Chiao Tung University } \\
{\it Hsinchu 30050, Taiwan, R. O. C.}}
\date{\today}
\maketitle

\begin{abstract}
A systematic calculation of magnetization and specific heat contributions
due to fluctuations of vortex lattice in strongly type II superconductors to
precision of 1\% is presented. We complete the calculation of the two loop
low temperature perturbation theory by including the umklapp processes. Then
the gaussian variational method is adapted to calculation of thermodynamic
characteristics of the 2D and the 3D vortex\ solids in high magnetic field.
Based on it as a starting point for a perturbation theory we calculate the
leading correction providing simultaneously an estimate of precision. The
results are compared to existing \ nonperturbative approaches.
\end{abstract}


\newpage

\section{Introduction}

Existence of vortex lattice in type II superconductors in magnetic field was
predicted by Abrikosov and subsequently observed in various types of such
superconductors ranging from metals to high $T_{c}$ cuprates. In the
original treatment mean field Ginzburg - Landau (GL) theory which neglects
thermal fluctuations of the vortex matter was used. Thermal fluctuations are
expected to play much larger role in high $T_{c}$ superconductors than in
the low temperature ones because the Ginzburg parameter $Gi$ characterizing
fluctuations is much larger \cite{Blatter}. In addition the presence of
strong magnetic field and strong anisotropy in superconductors like BSCCO
effectively reduces their dimensionality thereby further enhancing effects
of thermal fluctuations. Under these circumstances fluctuations make the
lattice softer influencing various physical properties like magnetization
and specific heat and eventually lead to vortex lattice melting into vortex
liquid far below the mean field phase transition line \cite{Nelson,Blatter}
clearly seen in both magnetization \cite{Zeldov} and specific heat
experiments \cite{Schilling}. To develop a theory of these fluctuations even
in the case of lowest Landau level (LLL) corresponding to regions of the
phase diagram ''close'' to $H_{c2}$ is a nontrivial task and several
different approaches were developed.

At high temperature end (and thereby in the ''vortex liquid'' phase)
Thouless and Ruggeri \cite{Ruggeri,Ruggeri1} proposed a perturbative
expansion around a homogeneous (liquid) state in which all the ''bubble''
diagrams are resummed. It was shown in field theory that summation of bubble
diagrams is equivalent to the gaussian variational approach \cite{Barnes}.
In this approach one searches for a ''gaussian'' state having the lowest
energy. The series provides accurate results at high temperatures, but
become inapplicable for LLL dimensionless temperature $a_{T}$ $\thicksim $ $%
(T-T_{mf}(H))/(TH)^{1/2}$ smaller than $2$ in 2D and for $a_{T}$ $\thicksim $
$(T-T_{mf}(H))/(TH)^{2/3}$ smaller than $1$ in 3D both quite far above the
melting line. Generally, attempts to extend the theory to lower temperature
by Pade extrapolation were not successful \cite{Moore1}. It is in fact
doubtful whether the perturbative results based on gaussian approximation
assuming translational invariant liquid state should be attempted at low $%
a_{T}$.

In ref.\cite{LiPRL}, \ it was shown that below $a_{T}<-5$ different gaussian
states which are no longer translationally invariant have lower energy. We
will present the detailed calculation here (optimized perturbation theory
was used to study for liquid state, see refs. \cite{LiPRL,PRBLIQ}). It is in
general a very nontrivial problem to find an inhomogeneous solutions of the
corresponding ''gap equation'' (see section IV). However using previous
experience with low temperature perturbation theory \cite{Rosenstein,Kao},
the problem can be significantly simplified and solved using rapidly
convergent ''modes'' expansion. A consistent perturbation theory should
start from these states \cite{Mermin}. We then generalize the approach of
ref. \cite{Ruggeri} by setting up a perturbation theory around the gaussian
Abrikosov lattice state.

Magnetization and specific heat contributions due to vortex lattices are
calculated in perturbation theory around this state to next to leading
order. This allows to estimate the precision of the calculation. It is the
worst, about 1\%, near the melting point at $a_{T}=-10$ and becomes better
for lower $a_{T}$. At low temperature the result is consistent with the
first principles low temperature perturbation theory advanced recently to
the two loop order \cite{Rosenstein,Kao}. The previous two loop calculation
is completed by including the umklapp processes. One can make several
definitive qualitative conclusions using the improved accuracy of the
results. The LLL scaled specific heat monotonously rises from its mean field
value of $1/\beta _{A}$ at $a_{T}=-\infty $ to a slightly higher value of $%
1.05/\beta _{A}$ where $\beta _{A}=1.16$ is Abrikosov parameter. This is at
variance with theory ref.\cite{Tesanovic} which uses completely different
ideas and has a freedom of arbitrarily choosing certain parameters on a 2\%
precision level. Although we calculate the contribution of the LLL only,
corrections due to higher Landau levels calculated earlier in ref.\cite
{Li1,Li2} using less sophisticated method can be included.

The paper is organized as follows. The model is defined in section II. In
section III a brief summary of existing results and umklapp corrected two
loop perturbative results in both 2D and 3D are given. Gaussian
approximation and the mode expansion used is described in section IV. The
basic idea of expansion around the best gaussian state is explained in
section V. The leading corrections are calculated. Results are presented and
compared with perturbation theory and other theories in section VI together
with conclusions

\section{Models}

To describe fluctuations of order parameter in thin films or layered
superconductors one can start with the Ginzburg-Landau free energy: 
\begin{equation}
F=L_{z}\int d^{2}x\frac{{\hbar }^{2}}{2m_{ab}}\left| D\psi \right|
^{2}-a|\psi |^{2}+\frac{b^{\prime }}{2}|\psi |^{4},
\end{equation}
where ${\bf A}=(By,0)$ describes a constant magnetic field (considered
nonfluctuating) in Landau gauge and covariant derivative is defined by ${\bf %
D}\equiv {\bf \nabla }-i\frac{2\pi }{\Phi _{0}}{\bf A,}\Phi _{0}\equiv \frac{%
hc}{e^{\ast }}$. For strongly type II superconductors like the high $T_{c}$
cuprates ($\kappa \sim 100$) and not too far from $H_{c2}$ (this is the
range of interest in this paper, for the detailed discussion of the range of
applicability see \cite{Li1}) magnetic field is homogeneous to a high degree
due to superposition from many vortices. For simplicity we assume $%
a(T)=\alpha T_{c}(1-t)$, $t\equiv T/T_{c},$ although this temperature
dependence can be easily modified to better describe the experimental $%
H_{c2}(T)$. The thickness of a layer is $L_{z}$.

Throughout most of the paper will use the coherence length $\xi =\sqrt{{%
\hbar }^{2}/\left( 2m_{ab}\alpha T_{c}\right) }$ as a unit of length and $%
\frac{dH_{c2}(T_{c})}{dT}T_{c}=\frac{\Phi _{0}}{2\pi \xi ^{2}}$ as a unit of
magnetic field. After the order parameter field is rescaled as $\psi
^{2}\rightarrow \frac{2\alpha T_{c}}{b^{\prime }}\psi ^{2}$, the
dimensionless free energy (the Boltzmann factor) is: 
\begin{equation}
\frac{F}{T}=\frac{1}{\omega }\int d^{2}x\left[ \frac{1}{2}|{\bf D}\psi |^{2}-%
\frac{1-t}{2}|\psi |^{2}+\frac{1}{2}|\psi |^{4}\right] .  \label{energ1}
\end{equation}
The dimensionless coefficient describing the strength of fluctuations is 
\begin{equation}
\omega =\sqrt{2Gi}\pi ^{2}t=\frac{m_{ab}b^{\prime }}{2{\hbar }^{2}\alpha
L_{z}}t,\;Gi\equiv \frac{1}{2}\left( \frac{32\pi e^{2}\kappa ^{2}\xi
^{2}T_{c}}{c^{2}h^{2}L_{z}}\right) ^{2}  \label{omega}
\end{equation}
where $Gi$ is the Ginzburg number in 2D . When $\frac{1-t-b}{12b}<<1$, the
lowest Landau level (LLL) approximation can be used \cite{Li1}. The model
then simplifies due to the LLL constraint, $-\frac{{\bf D}^{2}}{2}\psi =%
\frac{b}{2}\psi $, rescaling $x\rightarrow x/\sqrt{b},y\rightarrow y/\sqrt{b}
$ and $|\psi |^{2}\rightarrow |\psi |^{2}\sqrt{\frac{b\omega }{4\pi }},$ one
obtains 
\begin{equation}
f=\frac{1}{4\pi }\int d^{2}x\left[ a_{T}|\psi |^{2}+\frac{1}{2}|\psi |^{4}%
\right] ,  \label{resca}
\end{equation}
where the 2D LLL reduced temperature 
\begin{equation}
a_{T}\equiv -\sqrt{\frac{4\pi }{b\omega }}\frac{1-t-b}{2}  \label{aT2D}
\end{equation}
is the only parameter in the theory \cite{Thouless,Ruggeri}.

For 3D materials with asymmetry along the $z$ axis the GL model takes a
form: 
\begin{equation}
F=\int d^{3}x\frac{{\hbar }^{2}}{2m_{ab}}\left| \left( {\bf \nabla }-\frac{%
ie^{\ast }}{\hbar c}{\bf A}\right) \psi \right| ^{2}+\frac{{\hbar }^{2}}{%
2m_{c}}|\partial _{z}\psi |^{2}+a|\psi |^{2}+\frac{b^{\prime }}{2}|\psi |^{4}
\end{equation}
which can be again rescaled into 
\begin{equation}
f=\frac{F}{T}=\frac{1}{\omega }\int d^{3}x\left[ \frac{1}{2}|{\bf D}\psi
|^{2}+\frac{1}{2}|\partial _{z}\psi |^{2}-\frac{1-t}{2}|\psi |^{2}+\frac{1}{2%
}|\psi |^{4}\right] ,
\end{equation}
by $x\rightarrow \xi x,y\rightarrow \xi y,z\rightarrow \frac{\xi z}{\gamma
^{1/2}},\psi ^{2}\rightarrow \frac{2\alpha T_{c}}{b^{\prime }}\psi ^{2},$
where $\gamma \equiv m_{c}/m_{ab}$ is anisotropy. The Ginzburg number is now
given by 
\begin{equation}
Gi\equiv \frac{1}{2}\left( \frac{32\pi e^{2}\kappa ^{2}\xi T_{c}\gamma ^{1/2}%
}{c^{2}h^{2}}\right) ^{2}
\end{equation}
Within the LLL approximation,and rescaling $x\rightarrow x/\sqrt{b}%
,y\rightarrow y/\sqrt{b},z\rightarrow z\left( \frac{b\omega }{4\pi \sqrt{2}}%
\right) ^{-1/3},\psi ^{2}\rightarrow \left( \frac{b\omega }{4\pi \sqrt{2}}%
\right) ^{2/3}\psi ^{2}$, the dimensionless free energy becomes: 
\begin{equation}
f=\frac{1}{4\pi \sqrt{2}}\int d^{3}x\left[ \frac{1}{2}|\partial _{z}\psi
|^{2}+a_{T}|\psi |^{2}+\frac{1}{2}|\psi |^{4}\right] .
\end{equation}
The 3D reduced temperature is: 
\begin{mathletters}
\begin{equation}
a_{T}=-\left( \frac{b\omega }{4\pi \sqrt{2}}\right) ^{-2/3}\frac{1-t-b}{2}.
\label{aT3D}
\end{equation}
From now on we work with rescaled quantities only and related them to
measured quantities in section V.

\section{Perturbation theories and existing nonperturbative results}

A variety of perturbative as well as nonperturbative methods have been used
to study this seemingly simple model. There are two phases. Neglecting
thermal fluctuations, one obtains the lowest energy configuration $\psi =0$
for $a_{T}>0$ and 
\end{mathletters}
\begin{eqnarray}
\psi &=&\sqrt{-\frac{a_{T}}{\beta _{A}}}\varphi ;  \nonumber \\
\varphi &=&\sqrt{\frac{2\pi }{\sqrt{\pi }a_{\bigtriangleup }}}%
\sum\limits_{l=-\infty }^{\infty }\exp \left\{ i\left[ \frac{\pi l(l-1)}{2}+%
\frac{2\pi }{a_{\bigtriangleup }}lx\right] -\frac{1}{2}(y-\frac{2\pi }{%
a_{\bigtriangleup }}l)^{2}\right\}  \label{phi}
\end{eqnarray}
for $a_{T}<0$, where $a_{\bigtriangleup }=\sqrt{\frac{4\pi }{\sqrt{3}}}$ is
the lattice spacing in our units and $\beta _{A}=1.16$.

\subsection{High temperature expansion in the liquid phase}

Homogeneous ''vortex liquid'' phase (which is not separated from the
''normal'' phase by a transition) has been studied using high temperature
perturbation theory by Thouless and Ruggeri \cite{Ruggeri}. Unfortunately
this asymptotic series (even pushed to a very high orders \cite{Hikami}) are
applicable only when $a_{T}>2.$ In order to extend the results to lower $%
a_{T}$ attempts have been made to Pade resum the series \cite{Ruggeri1}
imposing a constraint that the result matches the Abrikosov mean field as $%
a_{T}\rightarrow -\infty $, . However, if no matching to the limit of is
imposed, the perturbative results cannot be significantly improved \cite
{Moore1}. Experiments \cite{Zeldov,Schilling}, Monte Carlo simulations \cite
{MC} and nonperturbative Bragg chain approximation \cite{Zhuravlev} all
point out that there is a first order melting transition around $a_{T}=-12$.
If this is the case it is difficult to support such a constraint.

\subsection{Low temperature perturbation theory in the solid phase. Umklapp
processes}

Recently a low temperature perturbation theory around Abrikosov solution eq.
(\ref{phi}) was developed and shown to be consistent up to the two loop
order \cite{Rosenstein,Kao,Li2}. Since we will use in the present study the
same basis and notations and also will compare to the perturbative results
we recount here few basic expressions. The order parameter field $\psi $ is
divided into a nonfluctuating (mean field) part and a small fluctuation 
\begin{equation}
\psi (x)=\sqrt{\frac{-a_{T}}{\beta _{A}}}\varphi (x)+\chi (x).
\label{meaneq}
\end{equation}
The field $\chi $ can be expanded in a basis of quasi - momentum eigen
functions on LLL in 2D:

\begin{equation}
\varphi _{{\bf k}}=\sqrt{\frac{2\pi }{\sqrt{\pi }a_{\bigtriangleup }}}%
\sum\limits_{l=-\infty }^{\infty }\exp \left\{ i\left[ \frac{\pi l(l-1)}{2}+%
\frac{2\pi (x-k_{y})}{a_{\bigtriangleup }}l-xk_{x}\right] -\frac{1}{2}%
(y+k_{x}-\frac{2\pi }{a_{\bigtriangleup }}l)^{2}\right\} .
\end{equation}
Then we diagonalize the quadratic term of free energy eq. (\ref{resca}) to
obtain the spectrum. Instead of the complex field $\chi ,$ two ''real''
fields $O$ and $A$ will be used: 
\begin{eqnarray}
\chi (x) &=&\frac{1}{\sqrt{2}}\int_{{\bf k}}\frac{\exp [-i\theta
_{k}/2]\varphi _{{\bf k}}(x)}{\left( \sqrt{2\pi }\right) ^{2}}\left( O_{{\bf %
k}}+iA_{{\bf k}}\right)   \label{expan} \\
\chi ^{\ast }(x) &=&\frac{1}{\sqrt{2}}\int_{{\bf k}}\frac{\exp [i\theta
_{k}/2]\varphi _{{\bf k}}^{\ast }(x)}{\left( \sqrt{2\pi }\right) ^{2}}\left(
O_{-{\bf k}}-iA_{-{\bf k}}\right)   \nonumber
\end{eqnarray}
where $\gamma _{k}=|\gamma _{k}|\exp [i\theta _{k}]$ and definition of $%
\gamma _{k}$ (and all other definitions of functions) can be found in
Appendix A. The eigenvalues found by Eilenberger in ref.\cite{Eilenberger}
are 
\begin{eqnarray}
\epsilon _{A}({\bf k}) &=&-a_{T}\left( -1+\frac{2}{\beta _{A}}\beta _{k}-%
\frac{1}{\beta _{A}}|\gamma _{k}|\right)   \label{elspec} \\
\epsilon _{O}({\bf k}) &=&-a_{T}\left( -1+\frac{2}{\beta _{A}}\beta _{k}+%
\frac{1}{\beta _{A}}|\gamma _{k}|\right) .  \nonumber
\end{eqnarray}
where $\beta _{k},$is defined in Appendix A. In particular, when $%
k\rightarrow 0$ \cite{Maki} 
\begin{equation}
\epsilon _{A}\approx 0.12\left| a_{T}\right| |k|^{4}.
\end{equation}
The second excitation mode $\epsilon _{O}$ has a finite gap. The free energy
to the two loop order was calculated in \cite{Kao}, however the umklapp
processes were not included$\ $in some two loop corrections. These processes
correspond to momentum nonconserving (up to integer times inverse lattice
constant) four leg vertices ( see Appendix A eq. (\ref{umk})). We therefore
recalculated these coefficients. The result in 2D is (see Fig.1a,b): 
\begin{equation}
f_{eff2D}=-\frac{a_{T}^{2}}{2\beta _{A}}+2\log \frac{\left| a_{T}\right| }{%
4\pi ^{2}}-\frac{19.9}{a_{T}^{2}}+c_{v}.  \label{perbresl}
\end{equation}
where \ $c_{v}=\left\langle \log \left[ \left( \epsilon _{A}({\bf k}\right)
\epsilon _{O}({\bf k})/a_{T}^{2}\right] \right\rangle =-2.92$. In 3D, \
similar calculation (extending the one carried in ref.\cite{Rosenstein} to
umklapp processes) gives: 
\begin{equation}
f_{eff3D}=-\frac{a_{T}^{2}}{2\beta _{A}}+2.848\left| a_{T}\right| ^{1/2}+%
\frac{2.4}{a_{T}}.
\end{equation}

\subsection{Nonperturbative methods}

Few nonperturbative methods have been attempted. Tesanovic and coworkers 
\cite{Tesanovic} developed a method based on an approximate separation of
the two energy scales. The larger contribution (98\%) is the condensation
energy, while the smaller one (2\%) describes motion of the vortices. The
result for energy in 2D is: 
\begin{eqnarray}
f_{eff} &=&-\frac{a_{T}^{2}U^{2}}{4}-\frac{a_{T}^{2}U}{2}\sqrt{\frac{U^{2}}{4%
}+\frac{2}{a_{T}^{2}}}+2arc\sinh \left[ \frac{a_{T}U}{2\sqrt{2}}\right]
\label{teseq} \\
U &=&\frac{1}{2}\left[ \frac{1}{\sqrt{2}}+\frac{1}{\sqrt{\beta _{A}}}+\tanh %
\left[ \frac{a_{T}}{4\sqrt{2}}+\frac{1}{2}\right] \left( \frac{1}{\sqrt{2}}-%
\frac{1}{\sqrt{\beta _{A}}}\right) \right]  \nonumber
\end{eqnarray}
Corresponding expressions in 3D were also derived.

No\ melting phase transition is seen since it belongs to the 2\% which
cannot be accounted for within this approach. There exist several Monte
Carlo (MC)\ simulations of the system \cite{MC}. The expression eq. (\ref
{teseq}) agrees quite well with high temperature perturbation theory and MC\
simulations and has been used to fit both magnetization and specific heat
experiments \cite{Schilling}, but only to mean field level agrees with low
temperature perturbation theory. Expanding eq. (\ref{teseq}) in $1/a_{T},$
one obtains an opposite sign of the one loop contribution, see Fig.1a and
discussion in section VI.

Other interesting \ nonperturbative methods include the $1/N$ expansion \cite
{Kotlar,MooreN} and the ''Bragg chain fluctuation approximation'' \cite
{Zhuravlev}.

\section{Gaussian Variational Approach}

\subsection{General Anzatz}

Gaussian variational approach originated in quantum mechanics and has been
developed in various forms and areas of physics \cite{Kleinert,Cornwall}. In
quantum mechanics it consists of choosing a gaussian wave function which has
the lowest energy expectation value. When fermionic fields are present the
approximation corresponds to BCS or Hartree - Fock variational state. In
scalar field theory one optimizes the quadratic part of the free energy 
\begin{eqnarray}
f &=&\int -\frac{1}{2}\phi ^{a}D^{-1}\phi ^{a}+V(\phi ^{a})  \nonumber \\
&=&\int \frac{1}{2}\left( \phi ^{a}-v^{a}\right) G^{-1ab}\left( \phi
^{b}-v^{b}\right) +\widetilde{V}(G,\phi ^{a})  \label{simmodel} \\
&=&K+\widetilde{V}  \nonumber
\end{eqnarray}
To obtain ''shift'' $v^{a}$ and ''width of the gaussian'' $G$, one minimizes
the gaussian effective free energy \cite{Cornwall}, which is an exact upper
bound on the energy (see proof in \cite{Kleinert}). The result of the
gaussian approximation can be thought of as resummation of all the ''cacti''
or loop diagrams \cite{Ruggeri,Barnes}. Further corrections will be obtained
in section V by inserting this solution for $G$ and taking more terms in the
expansion of $Z$.

\subsection{2D Abrikosov vortex lattice}

In our case of one complex field one should consider the most general
quadratic form 
\begin{eqnarray}
K &=&\int_{x,y}\left( \psi ^{\ast }(x)-v^{\ast }(x)\right) G^{-1}(x,y)\left(
\psi (y)-v(y)\right) \\
&&+\left( \psi -v(x)\right) H^{\ast }\left( \psi -v(x)\right) +\left( \psi
^{\ast }-v^{\ast }(x)\right) H\left( \psi ^{\ast }-v^{\ast }(x)\right) 
\nonumber
\end{eqnarray}

Assuming hexagonal symmetry (a safe assumption for the present purpose), the
shift should be proportional to the mean field solution eq. (\ref{meaneq}), $%
v(x)=v\varphi (x),$ with a constant $v$ taken real thanks global $U(1)$
gauge symmetry. On LLL, as in perturbation theory, we will use variables $%
O_{k}$ and $A_{k}$ defined in eq. (\ref{expan}) instead of $\psi (x)$ 
\begin{equation}
\psi (x)=v\varphi (x)+\frac{1}{\sqrt{2}2\pi }\int_{{\bf k}}\exp \left[ -%
\frac{i\theta _{{\bf k}}}{2}\right] \varphi _{{\bf k}}(x)\left( O_{{\bf k}%
}+iA_{{\bf k}}\right) .  \label{expandgauss}
\end{equation}
The phase defined after eq. (\ref{expan}) is quite important for
simplification of the problem and was introduced for future convenience. The
most general quadratic form is 
\begin{eqnarray}
K &=&\frac{1}{8\pi }\int_{k}O_{{\bf k}}G_{OO}^{-1}(k)O_{-{\bf k}}+A_{{\bf k}%
}G_{AA}^{-1}(k)A_{-{\bf k}}+  \nonumber \\
&&O_{{\bf k}}G_{OA}^{-1}(k)A_{-{\bf k}}+A_{{\bf k}}G_{OA}^{-1}(k)O_{-{\bf k}%
},
\end{eqnarray}
with matrix of functions $G(k)$ on Brillouin zone to be determined together
with the constant$\ v$ by the variational principle. The corresponding
gaussian free energy is 
\begin{eqnarray}
f_{gauss} &=&a_{T}v^{2}+\frac{\beta _{A}}{2}v^{4}-2-\left\langle \log \left[
\left( 4\pi \right) ^{2}\det (G)\right] \right\rangle _{k}+  \nonumber \\
&&\left\langle a_{T}\left( G_{OO}\left( k\right) +G_{AA}\left( k\right)
\right) +v^{2}\left[ \left( 2\beta _{k}+\left| \gamma _{k}\right| \right)
G_{OO}\left( k\right) +\left( 2\beta _{k}-\left| \gamma _{k}\right| \right)
G_{AA}\left( k\right) \right] \right\rangle _{k}  \nonumber \\
&&\left\langle \beta _{k-l}\left[ G_{OO}\left( k\right) +G_{AA}\left(
k\right) \right] \left[ G_{OO}\left( l\right) +G_{AA}\left( l\right) \right]
\right\rangle _{k,l}+ \\
&&\frac{1}{2\beta _{A}}\left\{ \left\langle \left| \gamma _{k}\right| \left(
G_{OO}\left( k\right) -G_{AA}\left( k\right) \right) \right\rangle
^{2}+4\left\langle \left| \gamma _{k}\right| G_{OA}\left( k\right)
\right\rangle _{k}^{2}\right\}  \nonumber
\end{eqnarray}
where $\left\langle ...\right\rangle _{k}$ denotes average over Brillouin
zone. The minimization equations are: 
\begin{eqnarray}
v^{2} &=&-\frac{a_{T}}{\beta _{A}}-\frac{1}{\beta _{A}}\left\langle \left(
2\beta _{k}+\left| \gamma _{k}\right| \right) G_{OO}\left( k\right) +\left(
2\beta _{k}-\left| \gamma _{k}\right| \right) G_{AA}\left( k\right)
\right\rangle _{k}  \label{shift} \\
\left[ G(k)^{-1}\right] _{OO} &=&a_{T}+v^{2}\left( 2\beta _{k}+\left| \gamma
_{k}\right| \right) +  \label{OO} \\
&&\left\langle \left( 2\beta _{k-l}+\frac{\left| \gamma _{k}\right| \left|
\gamma _{l}\right| }{\beta _{A}}\right) G_{OO}\left( l\right) +\left( 2\beta
_{k-l}-\frac{\left| \gamma _{k}\right| \left| \gamma _{l}\right| }{\beta _{A}%
}\right) G_{AA}\left( k\right) \right\rangle _{l}  \nonumber \\
\left[ G(k)^{-1}\right] _{AA} &=&a_{T}+v^{2}\left( 2\beta _{k}-\left| \gamma
_{k}\right| \right) +  \label{AA} \\
&&\left\langle \left( 2\beta _{k-l}+\frac{\left| \gamma _{k}\right| \left|
\gamma _{l}\right| }{\beta _{A}}\right) G_{AA}\left( l\right) +\left( 2\beta
_{k-l}-\frac{\left| \gamma _{k}\right| \left| \gamma _{l}\right| }{\beta _{A}%
}\right) G_{OO}\left( k\right) \right\rangle _{l}  \nonumber \\
\left[ G(k)^{-1}\right] _{OA} &=&-\frac{G_{OA}(k)}{%
G_{OO}(k)G_{AA}(k)-G_{OA}(k)^{2}}=4\frac{\left| \gamma _{k}\right| }{\beta
_{A}}\left\langle \left| \gamma _{l}\right| G_{OA}\left( l\right)
\right\rangle _{l}  \label{OA}
\end{eqnarray}
These equations look quite intractable, however they can be simplified. The
crucial observation is that after we have inserted the phase $\exp [-i\theta
_{k}/2]$ in eq. (\ref{expandgauss}) using our experience with perturbation
theory, $G_{AO}$ appears explicitly only on the right hand side of the last
equation. It also implicitly appears on the left hand side due to a need to
invert the matrix $G$. Obviously $G_{OA}(k)=0$ is a solution and in this
case the matrix diagonalizes. However general solution can be shown to
differ from this simple one just by a global gauge transformation.
Subtracting eq. (\ref{OO}) from eq. (\ref{AA}) and using eq. (\ref{OA}), we
observe that matrix $G^{-1}$ has a form: 
\[
G^{-1}\equiv \left( 
\begin{array}{cc}
E_{O}(k) & E_{OA}(k) \\ 
E_{OA}(k) & E_{A}(k)
\end{array}
\right) =\left( 
\begin{array}{cc}
E(k)+\Delta _{1}\left| \gamma _{k}\right| & \Delta _{2}\left| \gamma
_{k}\right| \\ 
\Delta _{2}\left| \gamma _{k}\right| & E(k)-\Delta _{1}\left| \gamma
_{k}\right|
\end{array}
\right) 
\]
where $\Delta _{1},\Delta _{2}$ are constants. Substituting this into the
gaussian energy one finds that it depends on $\Delta _{1},\Delta _{2}$ via
the combination $\Delta =\sqrt{\Delta _{1}^{2}+\Delta _{2}^{2}}$ only.
Therefore without loss of generality we can set $\Delta _{2}=0$, thereby
returning to the $G_{OA}=0$ case \cite{footnote1}.

Using this observation the gap equations significantly simplify. The
function $E(k)$ and the constant $\Delta $ satisfy: 
\begin{eqnarray}
E(k) &=&a_{T}+2v^{2}\beta _{k}+2\left\langle \beta _{k-l}\left( \frac{1}{%
E_{O}(l)}+\frac{1}{E_{A}(l)}\right) \right\rangle _{l}  \label{mode} \\
\beta _{A}\Delta &=&a_{T}-2\left\langle \beta _{k}\left( \frac{1}{E_{O}(k)}+%
\frac{1}{E_{A}(k)}\right) \right\rangle _{k}.  \label{delta}
\end{eqnarray}
The gaussian energy becomes: 
\begin{eqnarray}
f &=&v^{2}a_{T}+\frac{\beta _{A}}{2}v^{4}+f_{1}+f_{2}+f_{3}  \nonumber \\
f_{1} &=&\left\langle \log \left[ \frac{E_{O}\left( k\right) }{4\pi ^{2}}%
\right] +\log \left[ \frac{E_{A}\left( k\right) }{4\pi ^{2}}\right]
\right\rangle _{k} \\
f_{2} &=&-2+\left\langle a_{T}\left( \frac{1}{E_{O}\left( k\right) }+\frac{1%
}{E_{A}\left( k\right) }\right) +v^{2}\left[ \left( 2\beta _{k}+\left|
\gamma _{k}\right| \right) \frac{1}{E_{O}\left( k\right) }+\left( 2\beta
_{k}-\left| \gamma _{k}\right| \right) \frac{1}{E_{A}\left( k\right) }\right]
\right\rangle _{k}  \nonumber \\
f_{3} &=&\left\langle \beta _{k-l}\left[ \frac{1}{E_{O}\left( k\right) }+%
\frac{1}{E_{A}\left( k\right) }\right] \left[ \frac{1}{E_{O}\left( l\right) }%
+\frac{1}{E_{A}\left( l\right) }\right] \right\rangle _{k,l}+  \nonumber \\
&&\frac{1}{2\beta _{A}}\left[ \left\langle \left| \gamma _{k}\right| \left( 
\frac{1}{E_{O}\left( k\right) }-\frac{1}{E_{A}\left( k\right) }\right)
\right\rangle _{k}\right] ^{2}
\end{eqnarray}

Using eq. (\ref{mode}), a formula 
\begin{eqnarray*}
\beta _{k} &=&\sum_{n=0}^{\infty }\chi ^{n}\beta _{n}(k) \\
\beta _{n}(k) &\equiv &\sum_{\left| {\bf X}\right| ^{2}=na_{\Delta
}^{2}}\exp [i{\bf k\bullet X}]
\end{eqnarray*}
derived in Appendix A and the hexagonal symmetry of the spectrum, one
deduces that $E(k)$ can be expanded in ''modes'' 
\begin{equation}
E(k)=\sum E_{n}\beta _{n}(k)
\end{equation}
The integer $n$ determines the distance of a points on reciprocal lattice
from the origin, see Fig. 4. and $\chi \equiv \exp [-a_{\Delta }^{2}/2]=\exp
[-2\pi /\sqrt{3}]=0.0265$. One estimates that $E_{n}\simeq \chi ^{n}a_{T},$
therefore the coefficients decrease exponentially with $n$. Note (see Fig.4)
that for some integers, for example $n=2,5,6$, $\beta _{n}=0$. Retaining
only first $s$ modes will be called ''the $s$ mode approximation''. We
miinimized numerically the gaussian energy by varying $v,\Delta $ and first
few modes of $E(k)$. The sample results for various $a_{T}$ and number of
modes are given in Table 1.

\begin{center}
{\bf Table 1.}

Mode expansion 2D.

\begin{tabular}{|c|c|c|c|}
\hline
$\ \ a_{T}$ & \ \ \ \ $1$ mode & \ \ $2$ modes & \ \ $3$ modes \\ \hline
$-1000$ & $-446023.8395$ & $-431171.9948$ & $-431171.9757$ \\ \hline
$-300$ & $-40131.29217$ & $-38796.0277$ & $-38796.02297$ \\ \hline
$-100$ & $-4450.41636$ & $-4303.28685$ & $-4303.28593$ \\ \hline
$-50$ & $-1106.51575$ & $-1070.63806$ & $-1070.63791$ \\ \hline
$-20$ & $-171.678045$ & $-166.690727$ & $-166.690827$ \\ \hline
$-10$ & $-39.292885$ & $-38.433571$ & $-38.433645$ \\ \hline
$-5$ & $-7.3153440$ & $-7.2237197$ & $-7.2237422$ \\ \hline
\end{tabular}
\end{center}

We see that in the interesting region of not very low temperatures the
energy converges extremely fast. In practice two modes are quite enough. The
results for the gaussian energy are plotted on Fig.1 and will be compared
with other approaches in section VI. Furthermore one can show that around $%
a_{T}<-4.6$, the gaussian liquid energy is larger than the gaussian solid
energy. So naturally when $a_{T}<-4.6$, one should use the gaussian solid to
set up a perturbation theory. For $a_{T}>-4.2$, there is no solution for the
gap equations.

\subsection{3D Abrikosov vortex lattice}

In \ 3D, we expand in bases of plan waves in the third direction times \
previously used quasi - momentum function: 
\begin{equation}
\psi ({\bf x},z)=v\varphi ({\bf x})+\frac{1}{\sqrt{2}\left( 2\pi \right)
^{3/2}}\int_{{\bf k,}k_{z}}\exp \left[ -\frac{i\theta _{{\bf k}}}{2}\right]
\varphi _{{\bf k}}({\bf x})\exp i\left( k_{z}\cdot z\right) \left(
O_{k}+iA_{k}\right) .
\end{equation}
. The quadratic form is 
\begin{equation}
K=\frac{1}{8\pi \sqrt{2}}%
\int_{k}O_{k}G_{OO}^{-1}(k)O_{-k}+A_{k}G_{AA}^{-1}(k)A_{-k}
\end{equation}
where integration over $k$ is understood as integration over Brillouin zone
and over $k_{z}$. Most of the derivation and important observations are
intact. The modifications are following 
\begin{eqnarray*}
G_{OO}^{-1}(k) &=&\frac{k_{z}^{2}}{2}+E_{O}({\bf k}) \\
G_{AA}^{-1}(k) &=&\frac{k_{z}^{2}}{2}+E_{A}({\bf k}).
\end{eqnarray*}

The corresponding gaussian free energy density (after integration over $%
k_{z} $) is: 
\begin{eqnarray}
f &=&v^{2}a_{T}+\frac{\beta _{A}}{2}v^{4}+f_{1}+f_{2}+f_{3}  \nonumber \\
f_{1} &=&\left\langle \sqrt{E_{O}({\bf k})}+\sqrt{E_{A}({\bf k})}%
\right\rangle _{{\bf k}} \\
f_{2} &=&a_{T}\left\langle \frac{1}{\sqrt{E_{O}({\bf k})}}+\frac{1}{\sqrt{%
E_{A}({\bf k})}}\right\rangle _{{\bf k}}+\left\langle v^{2}\left[ \left(
2\beta _{k}+\left| \gamma _{k}\right| \right) \frac{1}{\sqrt{E_{O}\left( 
{\bf k}\right) }}+\left( 2\beta _{k}-\left| \gamma _{k}\right| \right) \frac{%
1}{\sqrt{E_{A}\left( {\bf k}\right) }}\right] \right\rangle _{{\bf k}} 
\nonumber \\
f_{3} &=&\left\langle \beta _{k-l}\left[ \frac{1}{\sqrt{E_{O}({\bf k})}}+%
\frac{1}{\sqrt{E_{A}({\bf k})}}\right] \left[ \frac{1}{\sqrt{E_{O}({\bf l})}}%
+\frac{1}{\sqrt{E_{A}({\bf l})}}\right] \right\rangle _{k,l}+  \nonumber \\
&&\frac{1}{2\beta _{A}}\left[ \left\langle \left| \gamma _{k}\right| \left( 
\frac{1}{\sqrt{E_{O}({\bf k})}}-\frac{1}{\sqrt{E_{A}({\bf k})}}\right)
\right\rangle _{k}\right] ^{2}.
\end{eqnarray}
Minimizing the above energy, gap equations similar to that in 2D can be
obtained. One finds that 
\begin{eqnarray*}
E_{O}({\bf k}) &=&E({\bf k})+\Delta \left| \gamma _{k}\right| , \\
E_{A}({\bf k}) &=&E({\bf k})-\Delta \left| \gamma _{k}\right| .
\end{eqnarray*}
$E({\bf k})$ can be solved by modes expansion two. We minimized numerically
the gaussian energy by varying $v,\Delta $ and first few modes of $E(k)$.
The sample results of free energy density for various $a_{T}$ with $3$ modes
are given in Table 2.

\begin{center}
{\bf Table 2.}

Mode expansion 3D.

\begin{tabular}{|c|c|c|c|c|c|c|c|}
\hline
$a_{T}$ & $-300$ & $-100$ & $-50$ & $-30$ & $-20$ & $-10$ & $-5.5$ \\ \hline
$f$ & $-38757.2294$ & $-4283.2287$ & $-1057.6453$ & $-372.2690$ & $-159.5392$
& $-33.9873$ & $-6.5103$ \\ \hline
\end{tabular}
\end{center}

In practice two modes are also quite enough in 3D. As in the case of 2D, one
can show that around $a_{T}<-5.5$, the gaussian liquid energy is larger than
the gaussian solid energy. So naturally when $a_{T}<-5.5$, one should use
the gaussian solid to set up a perturbation theory in 3D. When around $%
a_{T}>-5$, there is no solution for the gap equations.

\section{Corrections to the gaussian approximation}

In this section, we calculate the lowest order correction to the gaussian
approximation (that will be called postgaussian correction), which will
determine the precision of the gaussian approximation. This is necessary in
order to fit experiments and compare with low temperature perturbation
theory and other nonperturbative methods.

First we review a general idea behind calculating systematic corrections to
the gaussian approximation \cite{Kleinert}. The procedure is rather similar
to calculating corrections to the Hartree-Fock approximations used in
fermionic systems. Gaussian variational principle provided us with the best
(in a certain sense) quadratic part of the free energy $K$ from which the
''steepest descent'' corrections can be calculated. The free energy is
divided into the quadratic part an a ''small'' perturbation $\widetilde{V}$.
For a general scalar theory defined in eq. (\ref{simmodel}) it takes a form: 
\begin{eqnarray}
f &=&K+\alpha \widetilde{V}  \nonumber \\
K &=&\frac{1}{2}\phi ^{a}G^{-1ab}\phi ^{b} \\
\widetilde{V} &=&-\frac{1}{2}\phi ^{a}D^{-1}\phi ^{a}+V(\phi ^{a})-\frac{1}{2%
}\phi ^{a}G^{-1ab}\phi ^{b}.  \nonumber
\end{eqnarray}
Here the auxiliary parameter $\alpha $ was introduced to set up a
perturbation theory. It will be set to one at the end of calculation.
Expanding logarithm of the statistical sum in powers of $\alpha $ 
\begin{equation}
Z=\int {\cal D}\phi ^{a}\exp (-K)\exp (-\alpha \widetilde{V})=\int {\cal D}%
\phi ^{a}\sum_{n=0}\frac{1}{n!}\left( \alpha \widetilde{V}\right) ^{n}\exp
(-K),
\end{equation}
one retains only first few terms. It was shown in refs. \cite{Cornwall} that
generally only two - particle irreducible diagrams contribute to the
postgaussian correction. The gaussian approximation corresponds to retaining
only first two terms, $n=0,1$, while the postgaussian correction retains in
addition the contribution of order $\alpha ^{2}$.

Feynman rules in our case are shown on Fig.5. We have two propagators for
fields $A$ and $O$ and three and four leg vertices. Using these rules the
postgaussian correction is presented on Fig. 6 as a set of two and three
loop diagrams. The corresponding expressions are given in Appendix B. The
Brillouin zone averages were computed numerically using the three modes
gaussian solution of the previous section. Now we turn to discussion of the
results.

\section{Results, comparison with other approaches and conclusion.}

Results for LLL scaled energy, magnetization and specific heat in 2D are
presented on Fig. 1, 2 and 3 respectively.

\subsection{Energy}

The gaussian energy provides a rigorous upper bound on free energy \cite
{Kleinert}. Fig.1a shows the 2D gaussian energy (the dash - dotted line),
which in the range of $a_{T}$ from $-30$ to $-10$ is just above the mean
field (the solid gray line). This is because it correctly accounts for the
(positive) logarithmic one loop correction of eq. (\ref{perbresl}). In
contrast the results of the theory by Tesanovic et al \cite{Tesanovic} (the
dashed gray line) are lower than the mean field. This reflects the fact that
although the correct large $\left| a_{T}\right| $\ limit is built in, the
expansion of the expression eq. (\ref{teseq}) gives negative coefficient of
the $\log \left| a_{T}\right| $ term. This is inconsistent with both the low
temperature perturbation theory and the gaussian approximation. The
difference between this theory and our result is smaller than 2\% only when $%
a_{T}<30$ or at small $a_{T}$ below the 2D melting line (which occurs at $%
a_{T}=-13$ according to Monte Carlo \cite{MC} and phenomenological estimates 
\cite{Hikami,Blatter}) where the lines become closer again. It never gets
larger than 10\% though. To effectively quantitatively study the model one
has to subtract the dominant mean field contribution. This is done in the
inset of Fig. 1a. We plot the gaussian result (the dash - dotted line), the
one loop perturbative result (the solid line) and eq. (\ref{teseq}) (the
dashed gray line) in an expanded region $-100<a_{T}<-10$. The gaussian
approximation is a bit higher than the one loop.

To determine the precision of the gaussian approximation and compare with
the perturbative two loop result, we further subtracted the one loop
contribution on Fig.1b. As expected the postgaussian result is lower than
the gaussian though higher than the two loop. The difference between the
gaussian and the postgaussian approximation in the region shown is about $%
\left| \Delta f\right| =0.2$, which translates into $0.2\%$ at $a_{T}=-30$,
0.4\% at $a_{T}=-20$ and 2\% at $a_{T}=-12$. The fit for the gauss and
postgaussian energy in the region $-30<a_{T}<-6$ are 
\begin{eqnarray*}
f_{g2D} &=&-\frac{a_{T}^{2}}{2\beta _{A}}+2\log \left| a_{T}\right| +0.119-%
\frac{19.104}{a_{T}}-\frac{60.527\log \left| a_{T}\right| }{a_{T}^{2}}+\frac{%
36.511}{a_{T}^{2}}+c_{v} \\
f_{pg2D} &=&-\frac{a_{T}^{2}}{2\beta _{A}}+2\log \left| a_{T}\right| +0.068-%
\frac{11.68}{a_{T}}-\frac{60.527\log \left| a_{T}\right| }{a_{T}^{2}}+\frac{%
38.705}{a_{T}^{2}}+c_{v}.
\end{eqnarray*}
In 3D, similarly one found that 
\begin{equation}
f_{g3D}=-\frac{a_{T}^{2}}{2\beta _{A}}+2.84835|a_{T}|^{1/2}+\frac{3.1777}{%
a_{T}}-\frac{0.8137\log ^{2}[-a_{T}]}{a_{T}}  \label{energ3d}
\end{equation}

\subsection{Magnetization}

The dimensionless LLL magnetization is defined as 
\begin{equation}
m(a_{T})=-\frac{df_{eff}(a_{T})}{da_{T}}  \label{magfor}
\end{equation}
and the measure magnetization is 
\begin{equation}
4\pi M=-\frac{e^{\ast }{h}}{cm_{ab}}\left\langle |\psi |^{2}\right\rangle =-%
\frac{e^{\ast }{h}}{cm_{ab}}|\psi _{r}|^{2}\frac{b^{\prime }}{2\alpha T_{c}}%
\sqrt{\frac{b\omega }{4\pi }},
\end{equation}
where $\psi $ is the order parameter of the original model, and $\psi _{r}$
is the rescaled one, which is equal to $\frac{df_{eff}(a_{T})}{da_{T}}$.
Thus 
\begin{equation}
4\pi M=\frac{e^{\ast }{h}}{cm_{ab}}\frac{b^{\prime }}{2\alpha T_{c}}\sqrt{%
\frac{b\omega }{4\pi }}m(a_{T}).  \label{magde}
\end{equation}
We plot the scaled magnetization in region $-30<a_{T}<-6$. Again, the mean
field contribution dominates, so we subtract it in Fig.2. The solid line is
the one loop approximation, while the gray line is the two loop
approximation. At small negative $a_{T}$ \ the postgaussian (the upper gray
dash - dotted line) is very close to the two loop result, while the gaussian
approximation (the dash - dotted line) is a bit lower. All of these lines
are above mean field. On the other hand, the result of \ ref.\cite{Tesanovic}
(the gray dashed line) is below the mean field. Magnetization jump at the
melting point is smaller than our precision of 2\% at $a_{T}=-12$. Our
result for the gaussian magnetization and the postgaussian correction in
this range can be conveniently fitted with 
\begin{eqnarray*}
m_{g2D} &=&\frac{a_{T}}{\beta _{A}}-\frac{2}{a_{T}}-\frac{19.10}{a_{T}^{2}}+%
\frac{133.55}{a_{T}^{3}}-\frac{121.05\log [-a_{T}]}{a_{T}^{3}} \\
\Delta m_{pg2D} &=&\frac{7.525}{a_{T}^{2}}-\frac{59.15}{a_{T}^{3}}+\frac{%
43.64\log [-a_{T}]}{a_{T}^{3}}
\end{eqnarray*}
respectively.

Similar discussion for the case of 3D can be deduced from eqs. (\ref{magfor}%
) , eq. (\ref{energ3d}) and 
\begin{eqnarray}
4\pi M &=&-\frac{e^{\ast }{h}}{cm_{ab}}\left\langle |\psi |^{2}\right\rangle
=-\frac{e^{\ast }{h}}{cm_{ab}}|\psi _{r}|^{2}\frac{b^{\prime }}{2\alpha T_{c}%
}\left( \frac{b\omega }{4\pi \sqrt{2}}\right) ^{2/3} \\
&=&\frac{e^{\ast }{h}}{cm_{ab}}\frac{b^{\prime }}{2\alpha T_{c}}\left( \frac{%
b\omega }{4\pi \sqrt{2}}\right) ^{2/3}m(a_{T}),  \nonumber
\end{eqnarray}
where the gaussian scaled magnetization can be obtained by differentiation
of eq. (\ref{energ3d}). We didn't attempt to calculate the postgaussian
correction in 3D.

\subsection{Specific heat}

The scaled LLL specific heat is defined as 
\begin{equation}
c(a_{T})=-\frac{d^{2}f_{eff}(a_{T})}{da_{T}^{2}}  \label{scaledc}
\end{equation}
and the original specific heat is related to the scaled specific heat $c$ in
2D\ via 
\[
C=\frac{1}{4\pi \xi ^{2}T}\left[ -b+\sqrt{\frac{\pi {\hbar }^{2}\alpha bT_{c}%
}{2m^{\ast }b^{\prime }T}}\frac{-3t-1+b}{2}m\left( a_{T}\right) +\frac{\pi {%
\hbar }^{2}\alpha T_{c}}{m^{\ast }b^{\prime }T}\frac{\left( -t-1+b\right) }{2%
}^{2}c(a_{T})\right] 
\]
We plot the scaled specific heat divided by the mean field value $%
c_{mf}=1/\beta _{A}$ in the range $-30<a_{T}<-6$ on Fig.3. The solid line is
one loop approximation, while the gray line is the two loop approximation.
At large $\left| a_{T}\right| $ \ the postgaussian (gray dashed - dotted
line) is very closed to the one loop result. Finally the gaussian
approximation (dash - dotted line) is a bit lower. All these lines are
slightly above mean field. On the contrary, the result of \ ref.\cite
{Tesanovic} (dashed gray line) is below the mean field. Our gaussian result
and its correction in this range can be conveniently fitted with: 
\begin{eqnarray*}
\frac{c_{g}}{c_{mf}} &=&1+\beta _{A}\left( \frac{2}{a_{T}^{2}}+\frac{38.2}{%
a_{T}^{3}}-\frac{521.7}{a_{T}^{4}}+\frac{363.2\ln [-a_{T}]}{a_{T}^{4}}\right)
\\
\frac{\Delta c_{g}}{c_{mf}} &=&\beta _{A}\left( -\frac{15.05}{a_{T}^{3}}+%
\frac{221.1}{a_{T}^{4}}-\frac{130.9\ln [-a_{T}]}{a_{T}^{4}}\right) .
\end{eqnarray*}
Qualitatively the gaussian specific heat is consistent with experiments \cite
{Schilling} which show that the specific heat first raise before dropping
sharply beyond the melting point.

\subsection{Conclusions}

In this paper, we applied the gaussian variational principle to the problem
of thermal fluctuations in vortex lattice state. Then the correction to it
was calculated perturbatively. This generalizes corresponding treatment of
fluctuations in the homogeneous phase (vortex liquid) by Thouless and
coworkers \cite{Ruggeri}. Also umklapp processes were included in the low
temperature two loop perturbation theory expression. The results of gaussian
perturbative and some nonperturbative approaches were compared. We hope that
increased sensitivity of both magnetization and specific heat experiments
will test the precision of the theory.

\acknowledgments
We are grateful to our colleagues A. Knigavko and T.K. Lee, B. Ya. Shapiro
and Y. Yeshurun for numerous discussions and encouragement. The work was
supported by NSC of Taiwan grant NSC$\#$89-2112-M-009-039

\appendix

\section{}

In this Appendix the basic definitions are collected. Brillouin zone
averages of products of four quasi - momentum functions are defined by: 
\begin{eqnarray}
\beta _{k} &=&<|\varphi |^{2}\varphi _{\overrightarrow{k}}\varphi _{%
\overrightarrow{k}}^{\ast }>  \nonumber \\
\gamma _{k} &=&<(\varphi ^{\ast })^{2}\varphi _{-\overrightarrow{k}}\varphi
_{\overrightarrow{k}}>  \label{beta} \\
\gamma _{k,l} &=&<\varphi _{k}^{\ast }\varphi _{-k}^{\ast }\varphi _{-%
\overrightarrow{l}}\varphi _{\overrightarrow{l}}>.  \nonumber
\end{eqnarray}
We also need a more general product $\left\langle \varphi _{k_{1}}^{\ast
}\varphi _{k_{2}}\varphi _{k_{3}}^{\ast }\varphi _{k_{4}}\right\rangle $ in
order to calculate postgaussian corrections. This is just a perturbative
four - leg vertex: 
\begin{eqnarray}
\left\langle \varphi _{k_{1}}^{\ast }\varphi _{k_{2}}\varphi _{k_{3}}^{\ast
}\varphi _{k_{4}}\right\rangle &=&\exp \left[ \frac{i\pi ^{2}}{2}\left(
n_{1}^{2}-n_{1}\right) +i\frac{2\pi }{a_{\Delta }}n_{1}k_{3y}\right] \delta
^{q}[{\bf k}_{1}-{\bf k}_{2}+{\bf k}_{3}-{\bf k}_{4}]\lambda \left[ {\bf k}%
_{1}-{\bf k}_{2},{\bf k}_{2}-{\bf k}_{4}\right] ,  \nonumber \\
\lambda \left[ {\bf l}_{1},{\bf l}_{2}\right] &=&\sum_{{\bf Q}}\exp [-\frac{%
\left| {\bf l}_{1}+{\bf Q}\right| ^{2}}{2}+i\left( l_{1x}+l_{2x}\right)
Q_{y}-i\left( l_{1y}+l_{2y}\right) Q_{x}]  \label{umk} \\
&&\times \exp \left[ i\left( l_{1x}+l_{2x}\right) l_{1y}\right] ,  \nonumber
\end{eqnarray}
where are reciprical lattice vectors: 
\begin{equation}
{\bf Q}=m_{1}\widetilde{{\bf d}}_{1}+m_{2}\widetilde{{\bf d}}_{2}.
\end{equation}
Here ${\bf k}_{1}-{\bf k}_{2}+{\bf k}_{3}-{\bf k}_{4}=n_{1}\widetilde{{\bf d}%
}_{1}+n_{2}\widetilde{{\bf d}}_{2}$ is assumed and the basis of reciprocal
lattice is $\widetilde{{\bf d}}_{1}=\frac{2\pi }{a_{\Delta }}\left( 1,-\frac{%
1}{\sqrt{3}}\right) ;\;\widetilde{{\bf d}}_{2}=\left( 0,\frac{4\pi }{%
a_{\Delta }\sqrt{3}}\right) ,a_{\Delta }=\sqrt{\frac{4\pi }{\sqrt{3}}}.$ It
is dual to $\ $the lattice{\bf \ }${\bf e}_{1}=(a_{\Delta },0),{\bf d}%
_{2}=\left( \frac{a_{\Delta }}{2},\frac{a_{\Delta }\sqrt{3}}{2}\right) $.
When ${\bf k}_{1}-{\bf k}_{2}+{\bf k}_{3}-{\bf k}_{4}\neq n_{1}\widetilde{%
{\bf d}}_{1}+n_{2}\widetilde{{\bf d}}_{2}$ the quantity vanishes. The delta
function differs from the Kroneker: 
\[
\delta ^{q}[{\bf k}]=\sum_{{\bf Q}}\delta \lbrack {\bf k+Q}]. 
\]
From the above formula, one gets the following expansion of $\beta _{k}$: 
\begin{eqnarray}
\beta _{k} &=&\sum_{m_{1,}m_{2}}\exp [-\frac{\left| {\bf X}\right| ^{2}}{2}+i%
{\bf k\bullet X}] \\
&=&\sum_{n}\exp \left[ -\frac{a_{\Delta }^{2}}{2}n\right] \beta _{n}(k) 
\nonumber
\end{eqnarray}
where ${\bf X=}n_{1}{\bf d}_{1}+n_{2}{\bf d}_{2}.$

To simplify the minimization equations we used the following general
identity. Any sixfold ($D_{6}$) symmetric function $F(k)$ (namely a function
satisfying $F(k)$ $=F(k^{^{\prime }}),$ where $k,k^{^{\prime }}$ is related
by a $\frac{2\pi }{6}$ rotation) obeys: 
\begin{equation}
\int F(k)\gamma _{k}\gamma _{k,l}=\frac{\gamma _{l}}{\beta _{A}}\int
F(k)\left| \gamma _{k}\right| ^{2}.
\end{equation}
This can be seen by expanding $F$ in Fourier modes and symmetrizing.

\section{}

In this Appendix we specify Feynman rules and collect expressions for
diagrams. The solid line Fig.5a represents $O$ and the dashed line Fig.5b
represents $A$. Fig.5c is a vertex with three $O$. In the coordinate space,
it is $2v\left[ \varphi O\left( O^{+}\right) ^{2}+c.c.\right] $. And Fig.5c
is $-2iv\varphi ^{+}O^{2}A^{+}-4iv\varphi OO^{+}A^{+}+c.c$. Fig.5g is $\frac{%
1}{2}\left| O\left( x\right) \right| ^{4}$. Fig.5h is $2OO^{+}\left(
AO^{+}-OA^{+}\right) $. Fig.5i is $4OO^{+}AA^{+}-\left[ O^{2}\left(
A^{+}\right) ^{2}+c.c\right] $.

Other vertices, for example, formulas for diagrams Fig.5e, f, j and k, can
be obtained by substituting $O\rightarrow iA,A\rightarrow -iO$ from formulas
for diagrams Fig.5d, c, h and g respectively. The propagator in coordinate
space can be written as 
\begin{eqnarray}
\left\langle O^{+}(x)O^{+}(y)\right\rangle  &=&4\pi \int_{{\bf k}%
}E_{O}\left( k\right) \varphi _{k}^{\ast }(x)\varphi _{-k}^{\ast }(y)=4\pi
P_{O}^{+}(x,y),  \nonumber \\
\left\langle O(x)O(y)\right\rangle  &=&4\pi \int_{{\bf k}}E_{O}\left(
k\right) \varphi _{k}(x)\varphi _{-k}(y)=4\pi P_{O}^{-}(x,y),  \label{greenf}
\\
\left\langle O(x)O^{+}(y)\right\rangle  &=&4\pi \int_{{\bf k}}E_{O}\left(
k\right) \varphi _{k}(x)\varphi _{k}^{\ast }(y)=4\pi P_{O}(x,y)  \nonumber
\end{eqnarray}
Functions $P_{A}^{+}(x,y),P_{A}^{-}(x,y),P_{A}(x,y)$ can be defined
similarly.

One finds three loops contribution to free energy from two \ $OOOO$ vertex
contraction, see Fig.6a, $-\frac{1}{16\left( 2\pi \right) ^{5}}\int_{{\bf x}%
}\left\langle f_{oooo}\right\rangle _{{\bf y}}$: 
\begin{equation}
f_{oooo}=4\left\langle \left| P_{O}\right| ^{4}+\left| P_{O}^{+}\right|
^{4}+4\left| P_{O}P_{O}^{+}\right| ^{2}\right\rangle _{{\bf y}}.
\label{oooo}
\end{equation}
{\bf \ }Coordinates are not written explicitly since all of them are the
same $P_{O}(x,y)\ $etc.

The contribution from the diagrams Fig.6b is $-\frac{1}{16\left( 2\pi
\right) ^{5}}\int_{{\bf x}}\left\langle f_{oooa}\right\rangle _{{\bf y}},$%
and 
\begin{eqnarray}
f_{oooa} &=&\left| P_{O}\right| ^{2}\left(
-16P_{O}^{+}P_{A}^{-}+8P_{O}P_{A}^{\ast }\right) +  \label{oooa} \\
&&\left| P_{O}^{+}\right| ^{2}\left( -8P_{O}^{+}P_{A}^{-}+16P_{O}P_{A}^{\ast
}\right) +c.c.  \nonumber
\end{eqnarray}
The diagrams Fig.6c is $-\frac{1}{16\left( 2\pi \right) ^{5}}\int_{{\bf x}%
}\left\langle f_{ooaa}\right\rangle _{{\bf y}}$ and 
\begin{eqnarray}
f_{ooaa} &=&16\left( \left| P_{O}\right| ^{2}+\left| P_{O}^{+}\right|
^{2}\right) \left( \left| P_{A}\right| ^{2}+\left| P_{A}^{+}\right|
^{2}\right) +  \nonumber \\
&&4\left( \left[ P_{O}^{+}\right] ^{2}\left[ P_{A}^{-}\right] ^{2}+P_{O}^{2}%
\left[ P_{A}^{2}\right] ^{\ast }+c.c.\right)   \nonumber \\
&&-32\left( P_{O}^{-}P_{O}^{\ast }P_{A}^{+}P_{A}+c.c.\right)   \label{ooaa}
\end{eqnarray}
The diagrams Fig.6f is $-\frac{v^{2}}{16\left( 2\pi \right) ^{4}}\int_{{\bf x%
}}\left\langle f_{ooo}\right\rangle _{{\bf y}}$ and 
\begin{eqnarray}
f_{ooo} &=&\left| P_{O}\right| ^{2}\left( 16P_{O}^{+}\varphi \left( x\right)
\varphi \left( y\right) +8P_{O}^{\ast }\varphi \left( x\right) \varphi
^{\ast }\left( y\right) +c.c.\right)   \nonumber \\
&&+\left| P_{O}^{+}\right| ^{2}\left( 8P_{O}^{+}\varphi \left( x\right)
\varphi \left( y\right) +16P_{O}^{\ast }\varphi \left( x\right) \varphi
^{\ast }\left( y\right) +c.c.\right)   \label{ooo}
\end{eqnarray}
The diagrams Fig.6h is $-\frac{v^{2}}{16\left( 2\pi \right) ^{4}}\int_{{\bf x%
}}\left\langle f_{ooa}\right\rangle _{{\bf y}}$and 
\begin{eqnarray}
f_{ooa} &=&-8\left( P_{O}^{-}\right) ^{2}P_{A}^{+}\varphi ^{\ast }\left(
x\right) \varphi ^{\ast }\left( y\right) -16\left( \left| P_{O}\right|
^{2}+\left| P_{O}^{+}\right| ^{2}\right) \times   \nonumber \\
&&\left( P_{A}^{+}\varphi \left( x\right) \varphi \left( y\right)
-P_{A}\varphi ^{\ast }\left( x\right) \varphi \left( y\right) \right)
+8P_{O}^{2}P_{A}^{\ast }\varphi ^{\ast }\left( x\right) \varphi \left(
y\right)   \nonumber \\
&&-32P_{O}P_{O}^{-}\left[ P_{A}^{+}\varphi ^{\ast }\left( x\right) \varphi
\left( y\right) -P_{A}^{\ast }\varphi ^{\ast }\left( x\right) \varphi ^{\ast
}\left( y\right) \right]   \label{ooa} \\
&&+c.c  \nonumber
\end{eqnarray}
Other contributions, Fig.6e,d,i,g can be obtained by substituting $%
P_{O}\longleftrightarrow P_{A,}P_{A}^{+}\longleftrightarrow
-P_{O,}^{+}P_{A}^{-}\longleftrightarrow -P_{O}^{-}$ in eq. (\ref{oooo}), eq.
(\ref{oooa}), eq. (\ref{ooo}) and eq. (\ref{ooa}).

\newpage

\begin{center}
{\Huge Figure captions}
\end{center}


{\LARGE Fig. 1a}

Scaled free energy of vortex solid. From top to bottom, gaussian
approximation (dash dotted line), mean field (solid line), theory ref. \cite
{Tesanovic} (dashed line). Inset: corrections to mean field calculated using
(from top to bottom) gaussian (dash - dotted line), one loop perturbation
theory (dotted line) and theory ref. \cite{Tesanovic} (dashed line).

{\LARGE Fig. 1b}

More refined comparison of different approximations to free energy. Mean
field as well as the one loop perturbative contributions are subtracted.

{\LARGE Fig. 2}

Thermal fluctuations correction to magnetization of vortex solid. From top
to bottom, one and two loop perturbation theory (solid lines ''p1'' and
''p2'' respectively), gaussian and postgaussian approximationa (dash dotted
lines ''g'' and ''pg'' respectively), theory ref. \cite{Tesanovic} (dashed
line ''t'').

{\LARGE Fig. 3}

Scaled specific heat eq.(\ref{scaledc}) normalized by the mean field. One
and two loop perturbation theory (solid lines ''p1'' and ''p2''
respectively), gaussian and postgaussian approximationa (dash dotted lines
''g'' and ''pg'' respectively), theory ref. \cite{Tesanovic} (dashed line
''t'').

{\LARGE Fig. 4}

Reciprical hexagonal lattice points ${\bf X}$ belonging to three lowest
order ''stars'' in the mode expansion of $\beta _{k}$.

{\LARGE Fig. 5}

Feynman rules of the low temperature perturbation theory. The solid line (a)
denoted the O mode propagator, while the dashed line (b) denotes the A mode
propagator. Various three leg and four leg vertices are presented on (c-f)
and (g-k) respectively.

{\LARGE Fig. 6}

Contributions to the postgaussian correction to free energy.


\begin{references}
\bibitem{footnote1}  Note that similar simplification occurs in the large $N$
approach developed in ref.\cite{Kotlar}. Consequently their Anzatz is in
fact a rigorous solution to their gap equations.
\end{references}
\end{document}